# Nanodiamond in tellurite glass Part II: practical nanodiamond-doped fibers

Yinlan Ruan,[1] Hong Ji,[1] Brett C. Johnson,[2] Takeshi Ohshima,[3] Andrew D. Greentree,[4] Brant C. Gibson,[4] Tanya M. Monro[1] and Heike Ebendorff-Heidepriem[1*]

[1] *ARC Centre of Excellence in Nanoscale BioPhotonics, Institute of Photonics and Advanced Sensing, The University of Adelaide, Adelaide, SA 5005, Australia*
[2] *Centre for Quantum Computing and Communication Technology, School of Physics, University of Melbourne, Parkville, Victoria 3010, Australia (previously at [3])*
[3] *Japan Atomic Energy Agency, Takasaki, Gunma 370-1292, Japan)*
[4] *ARC Centre of Excellence in Nanoscale BioPhotonics, School of Applied Sciences, RMIT University, Melbourne, VIC 3001, Australia*
[*]*heike.ebendorff@adelaide.edu.au*

**Abstract:** Tellurite glass fibers with embedded nanodiamond are attractive materials for quantum photonics applications. Reducing the loss of these fibers in the 600-800 nm wavelength range of nanodiamond fluorescence is essential to exploit the unique properties of nanodiamond in the new hybrid material. The first part of this study reported the origin of loss in nanodiamond-doped glass and impact of glass fabrication conditions. Here, we report the fabrication of nanodiamond-doped tellurite fibers with significantly reduced loss in the visible through further understanding of the impact of glass fabrication conditions on the interaction of the glass melt with the embedded nanodiamond. We fabricated tellurite fibers containing nanodiamond in concentrations up to 0.7 ppm-weight, while reducing the loss by more than an order of magnitude down to 10 dB/m at 600-800 nm.

©2014 Optical Society of America

**OCIS codes:** (060.2280) Fiber design and fabrication; (160.2220) Defect-center materials; (160.2290) Fiber materials; (160.2750) Glass and other amorphous materials, (160.4236) Nanomaterials.

## References and links

1. S. Kühn, C. Hettich, C. Schmitt, J.-Ph. Poizat, and V. Sandoghdar, "Diamond colour centres as a nanoscopic light source for scanning near-field optical microscopy," J. Microsc. **202**, 2-6 (2001).
2. A. Beveratos, R. Brouri, T. Gacoin, A. Villing, J.-P. Poizat, and P. Grangier, "Single Photon Quantum Cryptography," Phys. Rev. Lett. **89**, article 187901 (2002).
3. W. Schmunka, M. Rodenberger, S. Peters, H. Hofer, and S. Kück, "Radiometric calibration of single photon detectors by a single photon source based on NV-centers in diamond", J. Modern Optics **58**, 1252-1259 (2011).
4. G. Balasubramanian, I. Y. Chan, R. Kolesov, M. Al-Hmoud, J. Tisler, C. Shin, C. Kim, A. Wojcik, P. R. Hemmer, A. Krueger, T. Hanke, A. Leitenstorfer, R. Bratschitsch, F. Jelezko, and J. Wrachtrup, "Nanoscale imaging magnetometry with diamond spins under ambient conditions," Nature **455**, 648–652 (2008)
5. J. M. Taylor, P. Cappellaro, L. Childress, L. Jiang, D. Budker, P. R. Hemmer, A. Yacoby, R. Walsworth, and M. D. Lukin, "High-sensitivity diamond magnetometer with nanoscale resolution," Nature Phys. **4**, 810-816 (2008).
6. F. Dolde, H. Fedder, M. W. Doherty, T. Nöbauer, F. Rempp, G. Balasubramanian, T. Wolf, F. Reinhard, L. C. L. Hollenberg, F. Jelezko, and J. Wrachtrup, "Electric-field sensing using single diamond spins," Nature Phys. **7**, 459-463 (2011).
7. J. H. Cole, L. C. L. Hollenberg, "Scanning quantum decoherence microscopy, Nanotechnology 20, 495401 (2009); L P McGuinness, L T Hall, A Stacey, D A Simpson, C D Hill, J H Cole, K Ganesan, B C Gibson, S Prawer, P Mulvaney, F Jelezko, J Wrachtrup, R E Scholten and L C L Hollenberg, "Ambient nanoscale sensing with single spins using quantum decoherence", New J. Phys. **15**, article 073042 (2013)
8. E. Ampem-Lassen, D. A. Simpson, B. C. Gibson, S. Trpkovski, F. M. Hossain, S. T. Huntington, K. Ganesan, L. C. L. Hollenberg, and S. Prawer, "Nano-manipulation of diamond-based single photon sources," Opt. Express **17**, 11287-11293 (2009).

9. T. van der Sar, E. C. Heeres, G. M. Dmochowski, G. de Lange, L. Robledo, T. H. Oosterkamp, and R. Hanson, "Nanopositioning of a diamond nanocrystal containing a single nitrogen-vacancy defect center," Appl. Phys. Lett. **94**, article 173104 (2009).
10. T. Schröder, A. W. Schell, G. Kewes, T. Aichele, and O. Benson, "Fiber-integrated diamond-based single photon source," Nano Lett. **11**, 198-202 (2010).
11. T. Schröder, M. Fujiwara, T. Noda, H.-Q. Zhao, O. Benson, and S. Takeuchi, "A nanodiamond-tapered fiber system with high single-mode coupling efficiency," Opt. Express **20**, 10490-10497 (2012).
12. M. R. Henderson, B. C. Gibson, H. Ebendorff-Heidepriem, K. Kuan, S. Afshar V., J. O. Orwa, I. Aharonovich, S. Tomljenovic-Hanic, A. D. Greentree, S. Prawer, and T. M. Monro, "Diamond in tellurite glass: A new medium for quantum information," Adv. Mater. **23**, 2806-2810 (2011).
13. H. Ebendorff-Heidepriem, Y. Ruan, H. Ji, A. D. Greentree, B. C. Gibson, and T. M. Monro, "Nanodiamond in tellurite glass Part I: origin of loss in ND-doped glass," submitted to Opt. Mater. Express.
14. S. Manning, H. Ebendorff-Heidepriem, and T. M. Monro, "Ternary tellurite glasses for the fabrication of nonlinear optical fibers," Opt. Mater. Express **2**, 140-152, (2012).
15. M. R. Oermann, H. Ebendorff-Heidepriem, Y. Li, T.-C. Foo, and T. M. Monro, "Index matching between passive and active tellurite glasses for use in microstructured fiber lasers: Erbium doped lanthanum-tellurite glass," Opt. Express. **17**, 15578-15784 (2009)
16. J. S. Wang, E. M. Vogel, and E. Snitzer, "Tellurite glass: new candidate for fiber devices," Opt. Mater. **3**, 187-203 (1994).
17. A. Mori, "Tellurite-based fibers and their applications to optical communication networks," J. Ceram. Soc. Japan **116**, 1040-1051 (2008).
18. H. Ebendorff-Heidepriem, K. Kuan, M.R. Oermann, K. Knight, T.M. Monro, "Extruded tellurite glass and fibers with low OH content for mid-infrared applications," Optical Materials Express **2**, 432-442, (2012).
19. N. Da, A. A. Enany, N. Granzow, M. A. Schmidt, P. St. J. Russell, L. Wondraczek, "Interfacial reactions between tellurite melts and silica during the production of microstructured optical devices," J. Non-Cryst. Solids **357**, 1558-1563 (2011).
20. H. Ebendorff-Heidepriem, and T. M. Monro, "Analysis of glass flow during extrusion of optical fiber preforms," Opt. Mater. Express **2**, 304-320, (2012).
21. R. Hanbury Brown and R. Q. Twiss, "Correlation between photons in two coherent beams of light," Nature **177**, 27-29 (1956).
22. http://omlc.org/calc/mie_calc.html

## 1. Introduction

Fluorescent nanodiamonds (NDs) are rapidly becoming important for photonic and sensing applications. Photonic applications include nanodiamonds for SNOM light sources [1], quantum key distribution [2] and metrology [3]. Sensing applications typically leverage the superb quantum properties of nanodiamond for magnetometry [4,5], electrometry [6] and quantum sensing [7].

Whilst fluorescent nanodiamond has undeniable promise, there is a real difficulty in scaling most of the existing integration methods towards practical technologies. The problems can be understood relatively simply. In all ND fabrication techniques, most of the NDs do not have the desired fluorescence properties for a given task, for example the desired fluorescence might be from a rarely found color center, or the ND might have too few, or two many color centers for a given task. 'Pick and place' techniques can overcome the low probability of finding desirable centers [8-10] but typically are laborious and require expert handling. Emission from centers is usually treated as being isotropic and hence poorly captured by confocal setups, which can suffer from lack of robustness and large total apparatus size. Alternative approaches such as dip coating [11] are suitable, although the long-term robustness of the ND to the surface is problematic, especially in scanning probe type experiments.

We have recently developed ND-doped tellurite glass, which was demonstrated to be an attractive material for quantum photonics [12,13]. In this hybrid material, the single-photon emitting NDs are completely embedded in the tellurite glass, allowing the single photon emission of the NDs to be directly coupled to the bound modes of a fiber or indeed any photonic structure made from this material. Such embedding significantly enhances device robustness and coupling efficiency between the single photon emission and fiber modes. The glass fabrication technology is ideally suited to technological applications, as hundreds of meters of fiber can be made in the same draw process. One could then envisage searching the fiber for the required centers.

Key to the survival of the NDs during the glass fabrication process reported in [13] was the separation of the glass fabrication into a batch melting step and a subsequent ND doping step, allowing the use of a lower temperature for doping ND into the glass.

The exploitation of this unique material has been hampered by the relatively high loss (>100 dB/m) of the ND-doped tellurite fibers in the visible spectral region, where the NDs show their emission [13]. To gain understanding of the causes of loss in the ND-doped glasses, in the first part of this study [13], we investigated the interaction of the NDs with the tellurite glasses melted in gold crucibles. We found that NDs act as a chemically reducing agent, reacting with higher valence species in the melt such as tellurium (IV) ions (constituent of the glass network), gold ions (dissolved in the melt via corrosion of the gold crucible) and oxygen (dissolved in the melt via interaction with the atmosphere). The reaction of the NDs with these species resulted in the formation of lower valence species such as gold nanoparticles (GNP) and reduced tellurium species as well as in the oxidation (burning) of the NDs. All these reactions consume the carbon atoms of the NDs, ultimately leading to their undesirable disappearance. The GNPs and reduced tellurium species in the fabricated fibers cause high loss in the visible.

The extent of the reactions of NDs with higher valence species in the tellurite melts was found to depend on the amount of NDs added to the glass melt, the gold ion concentration, the amount of available oxygen in the melt and the batch melting temperature. Preliminary melting trials demonstrated complete oxidation of the NDs for times greater than 30 min in the hot melt before casting. Therefore, we limited this time to 10-20 min (minimum time to achieve dispersion of the NDs in the melt). Higher batch melting temperature increased the gold ion dissolution, which enhanced formation of high-loss GNPs. A high surface-to-volume ratio for small glass volumes increased the amount of oxygen from the atmosphere being dissolved into the glass melt, leading to enhanced burning of the NDs while reducing the formation of GNPs and reduced tellurium species. As the ND doping concentration increased, the amount of GNPs and reduced tellurium species increased, resulting in higher loss. These conclusions were obtained mainly for glass melts being not sufficiently large for fiber fabrication.

In this paper, we investigate the impact of fabrication conditions on loss and ND survival for large glass melts and fibers made from these glasses. Further improving the fabrication conditions, for example, by using silica crucibles combined with a lower melting temperature tellurite glass composition, we achieved ND-doped tellurite glass fibers with relatively low loss while preserving functional NDs. We evaluate the potential of these fibers for single photon propagation along these fibers.

## 2. Glass and crucible selection

### 2.2. Glass selection

We prepared glasses using two different Zn-Na-tellurite compositions labelled TZNL and TZN (Table 1). TZNL glasses were investigated in the first part of this study [13]. Here, we first continued exploring the impact of different fabrication conditions and ND doping concentrations for large TZNL melts made from 150 g batches. During the investigations, it was found that lower melting temperatures than that possible with TZNL glass was advantageous. As the glass melting temperature usually scales with the glass transition temperature, we searched for a tellurite glass having a glass transition temperature lower than that of TZNL (315 $^{o}$C). Prior work in our group identified a La-free Na-Zn-tellurite glass that exhibited a significantly lower glass transition temperature (293 °C) while demonstrating a comparable crystallization stability and refractive index relative to TZNL [14,15]. This La-free tellurite glass is hereafter referred to as TZN glass. The potential suitability of TZN glass for low-loss fiber fabrication was demonstrated by the fabrication of high-quality glass billets using large batches of up to 300 g [14]. Such glass billet sizes are sufficiently large to fabricate microstructured fibers using the extrusion technique. The composition and properties of the TZNL and TZN glasses are compared in Table 1.

Table 1. Glass composition and properties of TZNL and TZN glasses

| glass type | composition (mol%) | density (g/cm$^3$) | $T_g$ (°C) | $\Delta T^a$ (°C) | refractive index at 1064nm |
|---|---|---|---|---|---|
| TZNL | 73 TeO$_2$ – 20 ZnO – 5 Na$_2$O – 2 La$_2$O$_3$ | 5.35 | 315 | 165 | 2.00 |
| TZN | 75 TeO$_2$ – 15 ZnO – 10 Na$_2$O | 5.15 | 293 | 171 | 1.98 |

$^a$ Crystallisation stability defined as difference between onset of glass crystallization and glass transition temperatures [14,15].

*2.2. Crucible material selection*

The TZNL and TZN large-melt glasses were initially melted in a gold crucible, the most widely used crucible material for the fabrication of low-loss tellurite fibers [16-18]. During the investigations, we found that the gold crucible corrosion and thus GNP formation was significantly reduced by decreasing the batch melting temperature $T_1$. However, even using the lowest possible $T_1$ and $t_1$ to completely melt all the raw materials, the gold crucible corrosion was still sufficiently high to form GNPs when doping large melts with ≥10 ppm ND (Section 5). Therefore, we investigated the use of silica as an alternative crucible material.

Melting of a 20 g TZNL batch in a silica crucible using a high $T_1$ of 900 °C (as used for the fabrication of the first low-loss undoped TZNL fibers [15,18]) resulted in a completely opaque glass. The corrosion of the silica crucible by the tellurite melt is consistent with the reported interfacial reactions between tellurite melt and silica [19]. Building on the finding that gold crucible corrosion decreases with decreasing $T_1$, we investigated the use of a lower batch melting temperature on silica crucible corrosion. For a 20 g melt at 800 °C melting temperature, a TZNL glass with several larger (≥1 mm) white particles was obtained. To minimize the reaction of tellurite glass with silica, we used the lowest possible $T_1$ of 720 °C for the next small TZNL melt (30 g batch). Before all glass raw materials were completely melted, we observed the formation of new white particles in the glass melt. The final glass contained many white particles visible with the naked eye. These results demonstrate that low-loss TZNL glass cannot be melted in a silica crucible.

Consistent with its lower glass transition temperature, we found that TZN glass could be melted at a lower temperature than TZNL. The lowest possible $T_1$ required to completely melt all raw materials was found to be 690 °C, 30 °C lower than TZNL. Given this reduction in melting temperature, we also tested melting of TZN glass at the lowest possible $T_1$ in a silica crucible. First, we melted 30 g TZN batch at 690 °C, resulting in a transparent glass with only a few small particles barely visible with the naked eye. Next, we melted 100 g TZN batch at 690 °C. The large melt contained only a few small particles barely visible with the naked eye. Based on this promising result, we fabricated an ND-doped tellurite glass using a silica crucible and 100 g batch and compared its properties with that of the ND-doped tellurite glasses melted in gold crucibles as described below.

## 3. Experimental approach

### 3.1. Glass fabrication

The glass samples were prepared in gold or silica crucibles from batches in the range of 100-150 g (Table 1). These samples are referred to as large-melt samples, whereas previously studied samples (considered here for comparison) made from 20 g batches are referred to as small-melt samples. The large-melt glasses were made using the same experimental procedures as for the glasses made in the first part of this study [13]. Briefly, the glasses were first melted at a temperature $T_1$ for a time $t_1$ until all raw materials were completely molten. Then the temperature was set to the doping temperature $T_2$, ND powder was added (or not, in case of undoped glasses) to the hot melt and the melt was placed back in the furnace at $T_2$ for a time $t_2$. Finally, the glass melt was cast into a preheated brass mold, annealed at about the glass transition temperature and slowly cooled down to room temperature. We used short $t_2$ of 10-20 min to minimize ND oxidation while allowing sufficient time for homogenization of the ND in the glass melt. The experimental details of glass melting are described in [13]. The ND doping concentration refers to the concentration by weight of the ND powder added to the glass melt.

Previously, we observed that the degree of ND oxidation depended on the surface-to-volume ratio of a glass melt in the crucible [13]. For the silica and gold crucibles used, the surface-to-volume ratio of a tellurite melt is 0.73 cm$^{-1}$ and 0.84 cm$^{-1}$ for a 100 g batch, and 0.54 cm$^{-1}$ and 0.59 cm$^{-1}$ for a 150 g melt, respectively. Hence, for the same batch size, the surface-to-volume ratio in a silica crucible is 0.9 of that in a gold crucible, and for the same crucible, the ratio of a 100 g melt is 1.4 times larger than that of a 150 g melt. This demonstrates that the surface-to-volume ratios for 100-150 g melts are comparable for both gold and silica crucibles and therefore no significant variation on ND oxidation is expected for 100-150 g batches melted in either crucible type. By contrast, the surface-to-volume ratio of small 20 g tellurite melts in silica and gold crucibles is significantly higher (2.18 cm$^{-1}$ and 3.35 cm$^{-1}$, respectively) compared with the large 100-150 g melts.

We fabricated three series of ND-doped and undoped large-melt glasses: (C) TZNL glasses melted in gold crucibles, (D) TZN glasses melted in gold crucibles, and (E) TZN glasses melted in silica crucibles. The melting conditions of these glasses are listed in Table 2. Note that glass series (A) and (B) are described in the first part of this study [13].

### 3.2. Nanodiamond types

We used two types of ND particles. The majority of the glasses were melted using commercially available ND with ~45 nm diameter (NaBond). The same batch of this un-irradiated ND powder was used for the first part of this study [13]. A few glasses were doped with irradiated ND to maximize the number of added NDs that can be seen in the confocal fluoresence microscope images. The number of vacancies within the ND was increased by irradiating the material with high energy (2 MeV) electrons from a Cockcroft-Walton accelerator at a fluence of $10^{18}$ electrons/cm$^2$. The ND material was subsequently annealed in a vacuum at 800 °C for 2 hours to induce vacancy diffusion and NV formation.

### 3.3. Fiber fabrication

The large glass melts were cast into cylindrical glass billets of 30 mm diameter and 20-30 mm height for fiber fabrication. Except fiber C4, all the fibers made were unstructured fibers. C4 was a suspended core fiber with a large core of ~30 μm diameter. Note that the loss of the unstructured and suspended core fibers made from undoped glass was found to be comparable, which indicates that the microstructure did not result in excess loss relative to the unstructured fiber. To fabricate the fibers from the cylindrical billets, we first extruded the glass billets into preforms of 10-12 mm diameter using the billet extrusion technique [20]. The billets were extruded through stainless steel dies at 354 °C (TZNL) or 321 °C (TZN) with a speed of 0.2 mm/min. The rods were annealed at $T_g$ in ambient atmosphere and drawn down to fibers of ~160 μm outer diameter using a fiber drawing tower. The tower furnace body was purged with nitrogen or a mixture of 70% nitrogen and 30% oxygen to avoid formation of reduced tellurium species on the fiber surface.

Table 2. Glass fabrication conditions

| Glass No. | glass type | melt weight (g) | crucible material | ND content (ppm$_w$) | $T_1$ (°C) | $t_1$ (h) | $T_2$ (°C) | $t_2$ (min) | oxygen content[a] (%) |
|---|---|---|---|---|---|---|---|---|---|
| C1 | TZNL | 100 | gold | 9 | 900 | 1.3 | 900 | 15 | 20 |
| C3 | TZNL | 150 | gold | 20 | 800 | 1.0 | 700 | 12 | 20 |
| C4 | TZNL | 150 | gold | 11 | 800 | 1.0 | 700 | 17 | 20 |
| C5 | TZNL | 150 | gold | 10 | 750 | 0.9 | 650 | 20 | 20 |
| C6 | TZNL | 150 | gold | 7 | 720 | 1.3 | 620 | 17 | 20 |
| C7 | TZNL | 150 | gold | 0 | 800 | 0.6 | 700 | 11 | 20 |
| D1 | TZN | 150 | gold | 16 | 690 | 0.9 | 570 | 10 | 50 |
| D2 | TZN | 150 | gold | 7 | 690 | 1.3 | 690 | 15 | 20 |
| D3 | TZN | 150 | gold | 14[b] | 690 | 1.1 | 690 | 10 | 100 |
| D4 | TZN | 150 | gold | 9[b] | 690 | 0.8 | 610 | 10 | 100 |
| D5 | TZN | 100 | gold | 0 | 690 | 0.7 | 610 | 10 | 100 |
| E2 | TZN | 100 | silica | 12[b] | 690 | 0.7 | 610 | 10 | 100 |
| E3 | TZN | 100 | silica | 0 | 690 | 0.7 | 610 | 10 | 100 |

[a] oxygen content in the atmosphere used for the second melting step at $T_2$
[b] irradiated ND

*3.4. Spectroscopic and microscopic measurements*

The optical absorption of the fibers was measured using the cutback technique, white light source and commercial optical spectrum analyzer ranging from 400-1700 nm.

Scanning confocal fluorescence microscopy was used to detect the presence and distribution of fluorescent emitters in the samples. For the unstructured fibers (all samples except C4), we imaged the cleaved endfaces of short fiber pieces (1-3 cm length). For the microstructured fiber sample C4, we could not image the fiber endface reliably due to the structure. Therefore, we imaged the core region of the endface of a polished slice (~3 mm thick) prepared from the extruded preform. An excitation power of 500 μW was used at a wavelength λ =532 nm. The confocal images were taken in a spectral window of 650-750 nm. The image size was 100 μm × 100 μm with focal depth of ~1 μm and pixel size of 300 nm. The images were taken by focusing 2-10 μm below the surface to ensure that the NDs imaged were below the tellurite surface. The number of emitters in an image was determined using software ImageJ. For some samples, the emission spectrum of selected emitters was measured in the spectral window of 550-850 nm. In addition, the presence of single-photon-emitting NV center(s) inside ND was tested using an in-house built Hanbury Brown and Twiss single photon antibunching system [21] collecting emission in the 650-750 nm wavelength range.

## 4. Glass billet quality: ND agglomeration and bubbles

Visual inspection of the ND doped billets relative to the undoped billets revealed the following two features. The ND-doped billets contained some aggregations of brown particles that were visible with the naked eye. We attribute these particles to ND agglomeration. This observation is consistent with the result that we found significant variation of emitter numbers in confocal microscopy images taken for several samples of the same fiber (Table 3), which we attribute to inhomogeneous distribution of NDs in the fiber. We found approximately the same variation of emitter numbers for samples of fibers made using different melting conditions and ND doping concentration as for samples of the same fiber, indicating that the inhomogeneous distribution of NDs had a larger impact on the emitter number of a sample compared to the change in the glass melting conditions or ND doping concentration.

The other distinct feature of the ND-doped billets was the significant increase in the number of bubbles as shown in Figure 1 for the photos of the undoped and ND-doped TZNL and TZN billets. The bubbles in the ND-doped billets are attributed to carbon mon/dioxide formed due to ND oxidation. In contrast to the small-melt ND-doped samples [13], the large-melt ND-doped samples showed significantly enhanced number of bubbles. The small surface-to-volume ratio of the large melts hindered removal of bubbles through the glass melt surface into the surrounding atmosphere. Quantitative analysis of the impact of ND doping concentration and $T_2$ is difficult since the glasses were melted using different $T_2$ and $T_2$ has two opposite effects. High $T_2$ favors ND oxidation resulting in more bubbles, however at the same time high $T_2$ facilitates bubble removal due to lower glass viscosity.

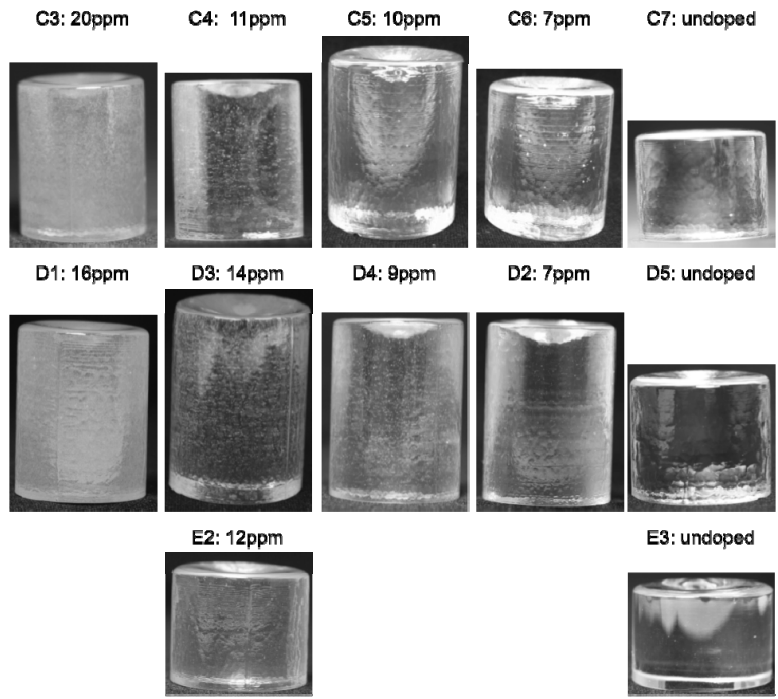

Fig. 1. Photographs of the glass billets fabricated, indicating amount of bubbles in the glass billets. The ND concentration decreases from left to right.

## 5. Impact of temperatures ($T_1$ and $T_2$) and ND doping concentration on fiber loss

We used the fiber loss spectra to investigate the impact of the glass fabrication conditions and ND doping concentration on the GNP formation. Figures 2a and c show the fiber loss spectra of the samples considered here. The majority of the samples show a broad absorption band at 600-700 nm. According to the previous results on small-melt glass samples [13], this band is attributed to the SPR absorption of GNPs in the fibers. To separate background loss from the SPR absorption, we used the following procedure. A background line going through the minimum loss at ~500 nm and the long wavelength tail of the SPR band at >700 nm was determined (Figs. 2a and c). The so-called SPR loss spectrum is then obtained by subtracting the background line from the measured loss spectrum. Figures 2b and d show the resultant SPR loss spectra of the fiber samples. We use the peak value of the SPR band as a measure of GNP content, whereas the background loss is affected by scattering from GNPs, agglomerated NDs and bubbles in the fibers as detailed towards the end of this section. Table 3 lists the values for SPR and background loss at the SPR peak wavelength for the fiber samples.

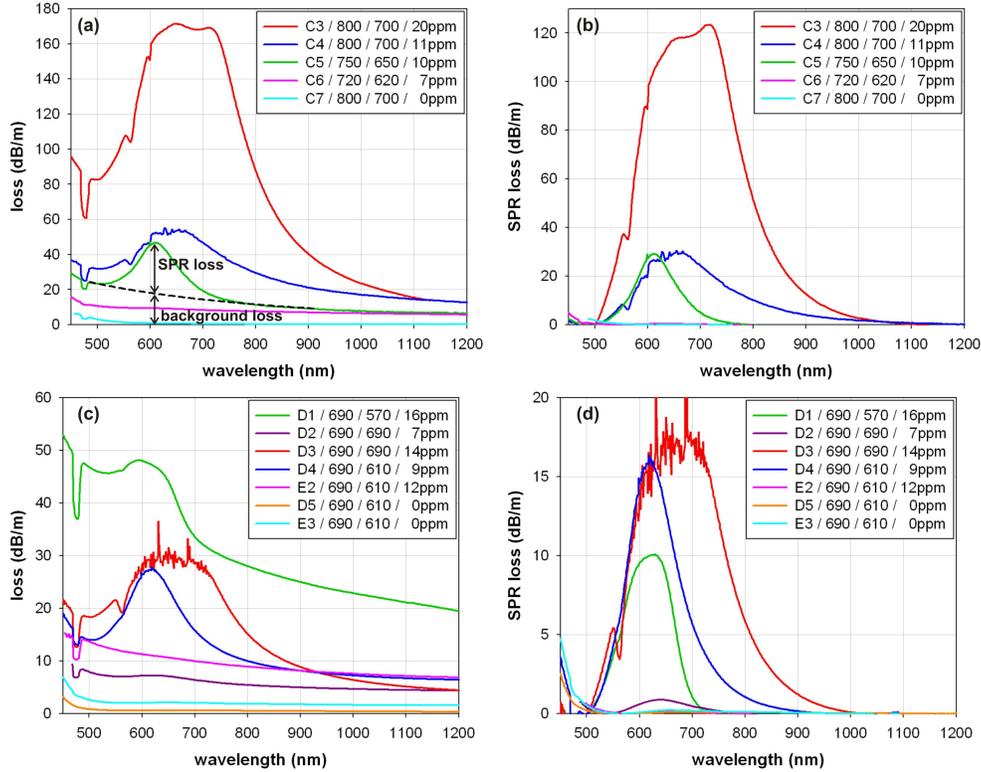

Fig. 2. Loss spectra of large-melt fiber samples. (a,b) TZNL samples and (c,d) TZN samples melted using different ND doping concentration and $T_1$ and $T_2$. (a,c) loss spectra as measured using the cutback technique, (b,d) SPR loss spectra calculated by subtracting the background loss.

We assume that the viscosity and time during extrusion and fiber drawing are in a regime where no reactions between oxygen, ND, tellurium species and gold species take place in the glass, and therefore the ND and GNP concentrations in preform and fiber samples are assumed to be the same as in the quenched glass melt (i.e. cast glass). This hypothesis is substantiated by the comparison of undoped TZNL fiber made from glasses melted at high and low $T_1$. For $T_1$=900 $^o$C, the corresponding fibers exhibited SPR peak absorption of ≤1 dB/m due to GNPs. By contrast, for low $T_1$=800 $^o$C, negligible SPR band was detected in corresponding fibers. The absence of significant SPR loss in fibers made from glasses melted and cast at 900 $^o$C indicates that, although gold ions were present in glass in high concentration due to high melting temperature, the extrusion conditions (<6 h at <360 $^o$C) and fiber drawing conditions (<30 min at ~400 $^o$C) were not sufficient to cause significant formation of GNPs.

Figure 3a shows SPR peak absorption in large-melt fiber samples (except C1, which was a preform sample, as the fiber sample could not be measured due to high loss) as a function of the temperature $T_1$. For comparison, the data for the small-melt samples of [13] are also shown. As for the small-melt samples, the SPR peak absorption of large-melt samples decreases with decreasing $T_1$, confirming that lower $T_1$ leads to lower GNP content due to reduced corrosion of the gold crucible.

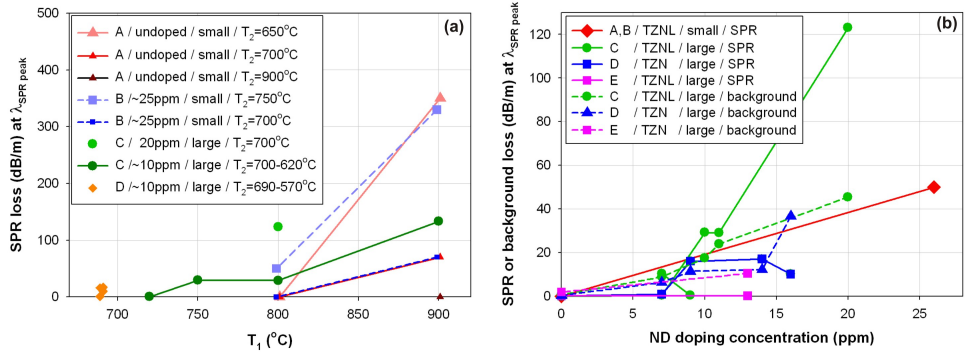

Fig. 3. (a) SPR peak absorption of undoped and ND-doped small-melt and large-melt samples melted in gold crucibles as a function of $T_1$ for different $T_2$ and (b) SPR and background absorption at the SPR peak wavelength as a function of ND doping concentration for small-melt and large-melt samples made using $T_1 \leq 800\ ^oC$ and $T_2 \leq 700\ ^oC$ as a function of ND doping concentration.

All large-melt fiber samples made using $T_1 < 750\ ^oC$ exhibit SPR peak absorption $\leq 17$ dB/m. For these samples, there is no clear correlation between SPR peak absorption and $T_1$. We assume that the GNP formation and therefore SPR peak absorption is sensitive to the various glass melting conditions used (temperature, time, atmosphere, gas flow, ND doping concentration). Consequently, for samples with low GNP content and thus low SPR peak absorption, the variation in SPR peak absorption caused by small variation in the fabrication conditions is larger than the impact of $T_1$ on SPR absorption.

Not surprisingly, the large-melt fiber sample E2 made using a silica crucible has negligible SPR peak absorption of 0.1 dB/m (Table 3, Fig. 3a) due to absence of gold crucible dissolution. The small SPR peak is attributed to gold impurities in the raw materials.

For small-melt TZNL samples made using the same $T_1/T_2$ temperatures, both background loss and SPR absorption increase with ND concentration [13]. For large-melt TZNL samples, SPR peak absorption also increases clearly with increasing ND concentration (Fig. 3b). These results for small-melt and large-melt TZNL samples confirm that ND doping enhances GNP formation.

For large-melt TZN samples, there is no clear correlation between SPR peak absorption and ND doping concentration. Since all the TZN glasses were melted at low $T_1$ of $<700\ ^oC$, they contained only small amount of gold ions and thus only small GNP amount and consequently low SPR absorption could be originated. As described above, for low SPR absorption, small changes in glass melting conditions mask the impact of a particular factor, e.g. ND concentration.

For higher ND doping concentration (>10 ppm), TZN glasses show lower fiber loss compared with TZNL, which is attributed to the lower melting temperature $T_1$ of TZN, resulting in reduced GNP formation in TZN.

Figure 3b shows that for all fiber samples, the background loss at the SPR peak wavelength generally increases with increasing ND doping concentration, which is attributed to enhanced bubble formation and ND agglomeration as the ND concentration increases. As a consequence, the total fiber loss (SPR and background) increases with increasing ND doping concentration. The behavior of SPR and background loss was also studied for constant wavelength of 690 nm, which is close to the peak wavelength of the ND fluorescence. The same trends as for the SPR peak wavelength were observed.

**Table 3. Properties of the fibers made**

| Sample No. | ND content (ppm) | SPR peak wavelength (nm) | SPR peak loss (dB/m) | background loss[a] (dB/m) | total number of emitters | AB dip found[b] | number of NV emitters | number of GNP emitters |
|---|---|---|---|---|---|---|---|---|
| C1 | 9 | 775 | 130[c] | n/a[d] | 140 | + | n/m | n/m |
| C3 | 20 | 715 | 123 | 54 | 170 | + | 11 | 10 |
| C4 | 11 | 658 | 29 | 24 | 7-14[c] | + | n/m | n/m |
| C5 | 10 | 612 | 29 | 16 | 94 | + | n/m | n/m |
| C6 | 7 | 629 | 0.4 | 9 | 6 | + | n/m | n/m |
| C7 | 0 | no peak | 0.0 | 0.7 | n/m | n/m | n/m | n/m |
| D1 | 16 | 628 | 10 | 35 | 75 | + | n/m | n/m |
| D2 | 7 | 639 | 0.9 | 6 | 27 | n/m | n/m | n/m |
| D3 | 14[e] | 671 | 17 | 13 | 50 | + | 20 | 0 |
| D4 | 9[e] | 619 | 16 | 11 | 1 | n/m | n/m | n/m |
| D5 | 0 | 650 | 0.1 | 0.4 | n/m | n/m | n/m | n/m |
| E2[f] | 12[e] | 655 | 0.1 | 10 | 230 | + | 6 | 0 |
| E3[f] | 0 | 656 | 0.2 | 1.9 | n/m | n/m | n/m | n/m |

[a] at SPR peak wavelength
[b] n/m is not measured, + and – designate whether emitter(s) with antibunching behavior were found (+) or not (–).
[c] measured for preform sample, whereas the other values of large melts are for fibre samples
[d] n/a is not applicable as the background loss value of the polished preform plate is not reliable due to surface imperfections of the sample as described in [13]
[e] irradiated ND
[f] glass was prepared using silica crucible, whereas other glasses were prepared using gold crucible

Figure 4 shows the background loss as a function of SPR absorption at 690 nm. Using Mie scattering theory [22], theoretical calculations of the SPR absorption and scattering of GNPs in high index medium (n=2.0) demonstrated that for GNPs the background loss as defined here (Fig. 2) is smaller than the SPR absorption around the peak wavelength. For many of the samples with low SPR absorption, the background loss is similar or higher than the SPR absorption, indicating that the fiber loss is not dominated by the SPR absorption or scattering of GNPs but by background loss due to other scattering defects. For ND-doped fibers with negligible SPR absorption, the impact of non-GNP related origins of loss could be investigated. The TZNL fibers C6 and C7 melted using gold crucibles have low SPR absorption of ≤0.4 dB/m; the background loss of fiber C6 doped with 7 ppm ND is considerably higher (9 dB/m) than the background loss of the undoped fiber C7 (0.7 dB/m). The TZN fibers E2 and E3 have also negligible SPR absorption due to melting in silica crucibles. Similarly to the TZNL fibers, the background loss of fiber E2 doped with 12 pm ND is considerably higher (10 dB/m) than the background loss of the undoped fiber E3 (1.9 dB/m). The excess loss in the ND-doped fibers relative to corresponding undoped fibers for both TZNL and TZN fibers is 8-9 dB/m. The higher loss of the ND-doped fibers is consistent with the presence of dark particles and large number of bubbles, resulting in strong scattering.

Comparison of the loss of undoped TZN fibers melted in gold and silica crucibles using the same melting conditions demonstrates that melting in a silica crucible increased the loss from 0.4 dB/m (D5 melted in gold crucible) to 1.9 dB/m (E3 melted in silica crucible). The higher loss of tellurite glasses melted in silica crucibles is attributed to the corrosion of silica by the hot melt, resulting in the formation of a few small white particles barely visible with the naked eye.

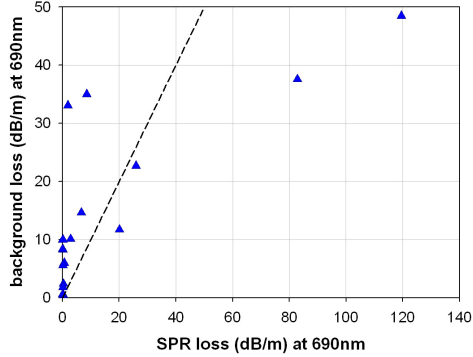

Fig. 4. Background loss as a function of SPR absorption at 690 nm. The triangles are the experimental background loss values for the fibers considered here and the black line designates where the background absorption equals the SPR absorption.

## 6. Confocal microscopy results: Detection of NDs in the tellurite fibers

We used scanning confocal microscopy to investigate ND survival in large-melt fiber samples. To distinguish between ND and GNP emitters in the confocal images, we measured antibunching behavior and emission spectra for selected samples. Previously, we found that GNPs and NVs exhibit different emission peak wavelengths. Emission spectra with peaks at ~650 nm are attributed to isolated GNPs, whereas emission spectra with peaks at ~690 nm are attributed to isolated NV centers in NDs [13]. We measured the emission spectra of several emitters in the fiber samples D3 and E2 (Fig. 5). For D3, the absence of GNP spectral behavior for any of the 20 emitters tested indicates that the low $T_1$ of 690 °C used for this glass melt suppressed the GNP formation. However, SPR peak absorption of 17 dB/m was found for the fiber sample, which indicates small amount of GNPs. The detection of SPR absorption but absence of GNP emitters demonstrates that for SPR peak absorption ≤17 dB/m of fiber loss spectra, the probability to find GNP emitters in confocal images is low. For E2, all of the 6 emitters tested also show only NV spectral behavior as was found for D3 (Fig. 5). This result is in agreement with the fabrication condition that E2 was melted in a silica crucible. In comparison, we found both GNP and NV emitters in C3 due to relatively high $T_1$ of 800 °C and high ND concentration of 20 ppm [13].

Another method to detect the presence of NDs in glass is the measurement of the antibunching behavior of an emitter. Figure 6a shows antibunching data from an emitter in the large-melt fiber sample C3, with the emitter exhibiting a characteristic dip at the zero delay time of the second order correlation $g^{(2)}$ of the emitter, which signifies antibunching of photons emitted from the emitter. This antibunching behavior is characteristic for a single-photon-emitting NV center inside ND. For the large-melt samples, the majority of the emitters tested did not show antibunching dip to zero but to significantly higher values (Fig. 6b). This behavior is attributed to enhanced background counts within or around the ND tested. The depth of an antibunching dip is directly proportional to the ratio of signal from the single-photon-emitting NV center to noise from other center(s)s within the ND or ND agglomeration, or other non-NV fluorescence within the same or neighboring NDs or the surrounding non-diamond medium within ~300 nm around the center. Hence, for the NV centers with antibunching dips >50% of $g^{(2)}(t)$ at $t=0$ ns, one of the above sources of noise is the reason for reduced depth of the antibunching dip. However, the presence of a dip at $t=0$ ns, irrespective of the depth of the dip, indicates the presence of one or more single-photon-emitting NV center(s) within the tested ND, and consequently that the glass contains active NDs exhibiting NV$^-$ centers.

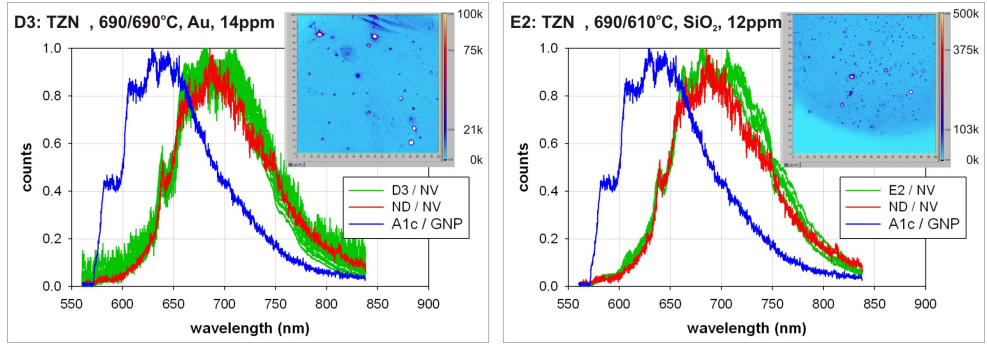

Fig. 5. Fluorescence spectra of emitters in D3 and E2 fiber samples (green lines), ND powder (red lines) and GNP in undoped A1c sample (blue lines) imaged using confocal microscopy. The spectra of the emitters in the fiber samples overlap with the spectrum of the emitters in ND powder, which indicates that all the emitters tested in the fiber samples are ND. The insets show the corresponding 100 μm × 100 μm confocal images of D3 and E2 samples.

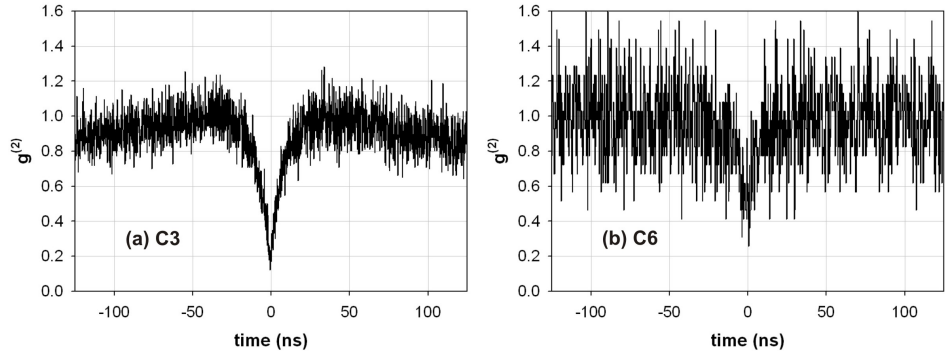

Fig. 6. Background corrected second order autocorrelation function of NV emitters in (a) C3 and (b) C6 fiber samples. Image (a) clearly shows the presence of a single NV center with $g^{(2)}(t)\sim0.1$. Image (b) likely indicates the presence of more than 1 NV center ($g^{(2)}(t)\sim0.5$) within the sampled ND.

Previously, we found that none of the emitters tested within the small-melt glass samples exhibited antibunching, whereas one large-melt fiber sample showed antibunching. In this work, all large-melt preform and fiber samples tested showed antibunching for at least some of the emitters (Table 3), confirming the presence of survived NDs in large-melt samples. In addition, the absence of a clear dependence of the number of emitters on the SPR absorption for large-melt samples indicates that the majority of emitters in the large-melt samples are not GNPs.

All three types of measurements (emitter spectra, antibunching behavior of emitters, number of emitters as a function of SPR absorption) indicate that some of the NDs added to a large glass melt survived, which we attribute to the significantly smaller surface-to-volume ratio of large melts compared with small melts, which reduced interaction of the melt with the atmosphere and thus with oxygen.

The amount of NDs that survived in a glass was estimated as follows. We first calculated the number of ND emitters, $N_{NDe}$, that is expected to be observed in the confocal volume, $V_{confocal}$, of an image assuming that all NDs doped into a melt survived,

$$N_{\text{NDe}} = w_{\text{ND}} \frac{V_{\text{confocal}} \cdot \rho_{\text{glass}}}{V_{\text{ND}} \cdot \rho_{\text{ND}}} x_e \qquad (1)$$

where $w_{\text{ND}}$ is the weight ratio of doped ND to glass melt, $\rho_{\text{glass}}$ and $\rho_{\text{ND}}$ are the density of the glass (5.35 g/cm$^3$ for TZNL and 5.15 g/cm$^3$ for TZN) and diamond (3.5 g/cm$^3$), respectively, $V_{\text{ND}}$ is the volume of an ND particle, and $x_e$ is the ratio of fluorescent ND (that are visible in a confocal image) to the total number of ND (that are present in the confocal volume). To calculate $V_{\text{ND}}$, we used 45 nm for the diameter of an ND particle. Based on the cross section of an image and assuming confocal depth of 1 μm, the confocal volume is estimated to be 100×100×1 μm$^3$. We assume that for irradiated ND all ND particles are fluorescent ($x_e$=1), whereas for unirradiated ND only 50% of the ND particles are fluorescent ($x_e$=0.5) [8]. Comparing the expected number of ND emitters calculated using Eq.1 with the number of ND emitters actually detected in an image yields the ratio of NDs that survived.

For the large-melt samples C3, D3 and E2, the ratio of fluorescent ND to the total number of emitters (ND and GNPs) is known from the spectral behavior of emitters tested. This ratio is ~60% for C3 and 100% for D3 and E2 as described above. Using Eq. 1 and the measured number of emitters (Table 3), the ratio of survived ND to doped ND is 3% for C3, 1% for D3 and 6% for E2. The corresponding concentrations of survived NDs are 0.2-0.7 ppm (Table 4). Note that these values of survived ND present a lower limit on the total survivability due to ND agglomeration. For the calculations, we assumed that one bright spot in a confocal image presents one ND particle of 45 nm diameter. However, the bright spots in the confocal images exhibit various sizes, indicating agglomeration of NDs. In addition, the diffraction limit of the confocal images is ~300 nm, which is larger than the size of one ND particle. Therefore, one bright spot in a confocal image could be caused by more than one ND particle, which would increase the portion of survived ND. In addition, the ND doping concentration that was determined from the weight of NDs added to the melt is a lower bound of the actual concentration in the glass volume due to accumulation on the melting crucible wall during addition of the ND powder to the melt.

Table 4. Rate and number of survived NDs in three fiber samples

| Sample No. | ND doping content (ppm) | ND survival rate (%) | ND survived content (ppm) | number of survived ND per 1m fiber of 160μm diameter (millions) | length of fiber with 160μm diameter containing 1 ND particle (nm) |
|---|---|---|---|---|---|
| C3 | 20 | 3.1 | 0.6 | 200 | 5 |
| D3 | 14[a] | 1.2 | 0.2 | 100 | 10 |
| E2 | 12[a] | 6.2 | 0.7 | 460 | 2 |

[a] irradiated ND

The determination of the amount of survived ND in the three fiber samples C3, D3 and E2 allowed us to investigate the impact of oxygen content in the melting atmosphere on the survival rate of ND. For the melting of the glasses of D3 and E2 100% oxygen was used and ND survival ratios of 1% and 6% were obtained, whereas the glass of C3 was melted using only 20% oxygen and ND survival ratio of 3% was obtained. Surprisingly, this result shows that the lower oxygen content for C3 did not result in a significant increase in the ratio of survived NDs. Therefore, we conclude that the amount of oxygen in the melting atmosphere has negligible impact on the ND survivability for ≥20% oxygen. This is an important result since higher oxygen concentration in the atmosphere is crucial to prevent formation of reduced tellurium species that lead to high loss in the visible.

The knowledge of the survival rate also allowed us to calculate the amount of survived NDs per fiber length assuming even distribution of the NDs in a fiber. By replacing the confocal volume in Eq. 1 with a certain fiber volume, Eq. 1 can be used to calculate the number of ND emitters expected to be found in this fiber volume. Using the survival rate, the number of survived NDs in this fiber volume can be calculated. Our unstructured fibers have 160 μm outer diameter, hence the number of survived NDs in 1 meter fiber length is 100-460 million, and the length of fiber to find one ND particle is 5-10 nm (Table 4). These values demonstrate that the amount of survived NDs in our fibers is more than sufficient to fabricate practical fiber lengths containing 1 single photon emitting ND. Lower ND concentration and improved dispersion of ND concentration is envisaged to result in single photon emitting fibers with loss of ~1 dB/m, enabling practical fiber lengths of ~1 meter.

## 7. Conclusions

We investigated the impact of a range of fabrication conditions on the loss of ND doped tellurite glass fibers and the ratio of survived NDs. GNPs and reduced tellurium species cause high loss in the visible due to SPR absorption at 600-700 nm and shift of the UV edge into the visible spectral region, respectively [13]. For tellurite glasses melted in a gold crucible, GNP formation decreases with decreasing batch melting temperature $T_1$ due to reduced gold crucible corrosion. ND doping facilitates the formation of GNPs due to chemically reducing effect of ND. Therefore, the GNP content increases with increasing ND doping concentration for melts with sufficient amount of gold ions to form GNPs in quantities that can be detected. For ND doping concentration >20 ppm, reduced tellurium species are formed, limiting the amount of NDs that can be doped into a glass while achieving low loss. Further experiments are required to determine the ND doping concentration limit with respect to reduced tellurium species for melts undertaken in a silica crucible. For high ND doping concentrations >10 ppm, TZN samples show lower GNP absorption than TZNL glasses due to lower $T_1$ used for TZN. For ND-doped glasses with low GNP content and absence of reduced tellurium species, the fiber loss is dominated by scattering from non-GNP particles (agglomerated NDs, bubbles, particles due to silica crucible dissolution). As the amount of bubbles is connected with the ND oxidation and thus with the ND doping concentration, and the background loss due to silica particles is fixed, the reduction of ND agglomeration is the key factor in further reducing the loss of ND-doped glasses and fibers in the future.

GNP emitters show shorter emission peak wavelength than NV emitters, allowing the type of emitters in our glasses to be identified and hence the amount of survived NDs. The ratio of survived NDs relative to doped ND amount was estimated to be in the range of a few percent for large-melt samples, whereas the amount of survived NDs in the small-melt samples was negligible. The higher ND survival probability for large melts is attributed to smaller surface-to-volume ratio, resulting in reduced interaction of the melt with oxygen in the atmosphere, which reduces the probability of oxidation of ND by dissolved oxygen in the melt. In contrast, the amount of oxygen in the melting atmosphere (20-100%) did not have a discernible effect on the ND survival probability, which suggests that the amount of oxygen in the melting atmosphere above the value that is needed to avoid tellurium ion reduction has no or weak influence on the ND oxidation for low $T_2$ temperatures of ≤700 °C used for the large-melt samples made in this work.

In conclusion, the optimum fabrication conditions for fibers with low-loss and large number of survived NDs are as follows. The batch melting temperature needs to be as low as possible to avoid dissolution of crucible material (gold ions, silica) into the glass melt. The ND doping temperature needs to be as low as possible to ensure survival of the NDs while being sufficiently high to ensure homogenization of the doped NDs in the glass liquid. The ND doping concentration is limited to ~20 ppm; higher concentration can lead to GNP and/or reduced tellurium species formation. The melt volume and thus surface-to-volume ratio needs to be in a range which allows sufficient oxygen in the melt to prevent GNP and reduced tellurium species formation, while limiting the oxygen in the melt to prevent complete oxidation of all NDs doped into the melt. Gold as a crucible material enables lower background loss but can enhance fiber loss due to SPR loss caused by GNP formation, in particular for larger melts and higher ND doping concentrations. In contrast, silica as crucible material results in higher background loss but, due to absence of GNP-related SPR loss, leads to lower fiber loss, in particular for higher ND doping concentrations. Based on these results, the optimum fabrication conditions to reduce loss while preserving ND in tellurite glass are to use a TZN glass composition enabling low $T_1=690\ ^oC$ and $T_2=610\ ^oC$ combined with use of silica crucible and ND doping concentrations <20 ppm. This finding is also of considerable importance for doping tellurite glass with non-diamond nanocrystals.

Using optimal fabrication conditions, future fibers with low loss and with active NV centers in NDs embedded in the glass will be used to investigate single photon emission and propagation along the fibers at room temperature. Such a system with possibly highest coupling efficiency will enable them to be used to connect various elements to integrate quantum information with the photonic backbones for quantum applications. In addition, these fibers will allow study of interaction of the NV centers with external magnetic field for potential applications in nanomagnetometry.

**Acknowledgements**

This work has been supported by ARC grants (DP120100901, DP130102494, FF0883189, FT110100225, LE100100104, CE110001027, CE140100003). T.M. acknowledges the support of an ARC Georgina Sweet Laureate Fellowship. This work was performed in part at the OptoFab node of the Australian National Fabrication Facility utilizing Commonwealth and SA State Government funding. We wish to thank Alastair Dowler and Tim Zhao at the University of Adelaide for fiber fabrication and help with measuring gold content in glass samples, respectively.